\documentclass[fleqn,usenatbib]{mnras}

\usepackage{newtxtext,newtxmath}
\usepackage[T1]{fontenc}
\usepackage{aecompl}

\usepackage{amsmath}
\usepackage{graphicx}
\usepackage{bm}

\title[Gravity anomaly from laboratory to astrophysics]{Gravity anomaly from laboratory experiments to astrophysics}

\author[Scarpa, Falomo, Treves]{
Riccardo Scarpa,$^{1}$
Renato Falomo,$^{2}$
Aldo Treves$^{3,4}$
\\
$^{1}$Instituto de Astrofisica de Canarias, C/O Via Lactea, s/n E-38205, La Laguna (Tenerife), Spain\\
$^{2}$INAF - Osservatorio Astronomico di Padova, vicolo dell'Osservatorio 5, I-35122, Padova, Italy\\
$^{3}$Universit\`a dell'Insubria, via Valeggio, 22100, Como, Italy\\
$^{4}$INAF - Osservatorio Astronomico di Brera, via Bianchi 46, I-23807, Merate (Lecco), Italy
}

\begin{document}

\maketitle

\begin{abstract}

Modifications to Newtonian dynamics at low accelerations have long been proposed as an alternative to dark matter to explain galaxy rotation curves. More recently, similar corrections have been invoked to interpret anomalies in Cavendish-type laboratory experiments and in the dynamics of wide binary stars, although the latter remain affected by ongoing observational debate.
We show that, if deviations from Newtonian gravity occur in the low-acceleration regime, the available data are broadly consistent with a MOND-like interpretation. In this framework, wide binary systems appear to extend the Tully–Fisher relation --well established for galaxies -- to much smaller mass scales.
Taken together, departures from Newtonian predictions at low accelerations in galaxies, wide binaries, and laboratory experiments may point to a common physical scenario.
\end{abstract}


\section{Introduction}
\label{intro}

The validity of Newtonian dynamics and of the law of gravitation, thoroughly tested within the Solar System, is often extrapolated over several orders of magnitude and applied to the extremely low-acceleration and weak-field regimes encountered in astrophysics. However, the legitimacy of such an extrapolation is not guaranteed.  
Early indications of possible departures from Newtonian predictions emerged from studies of galaxy clusters  and from the rotation curves of galaxies, which revealed significant discrepancies when the observed baryonic mass was used to predict the dynamical behavior (see, e.g., \citealt{Lelli17} for a review and  references). These discrepancies are commonly attributed to the presence of large amounts of non-baryonic dark matter (e.g., \citealt{Bertone}), although alternative explanations invoking revisions of the underlying gravitational laws have also been proposed. Among these, a specific modification of Newtonian dynamics (MOND) introduced by \citet{Milgrom}, posits a deviation from the standard Newtonian formulation below a characteristic acceleration scale $a_0 \sim 1.2 \times 10^{-10}\ \mathrm{m\ s^{-2}}$. This framework has attracted considerable attention owing to its empirical success in reproducing a wide range of galaxy rotation curve data (e.g., \citealt{Sanders}).

\begin{figure}
\centering
\includegraphics[width=1\columnwidth]{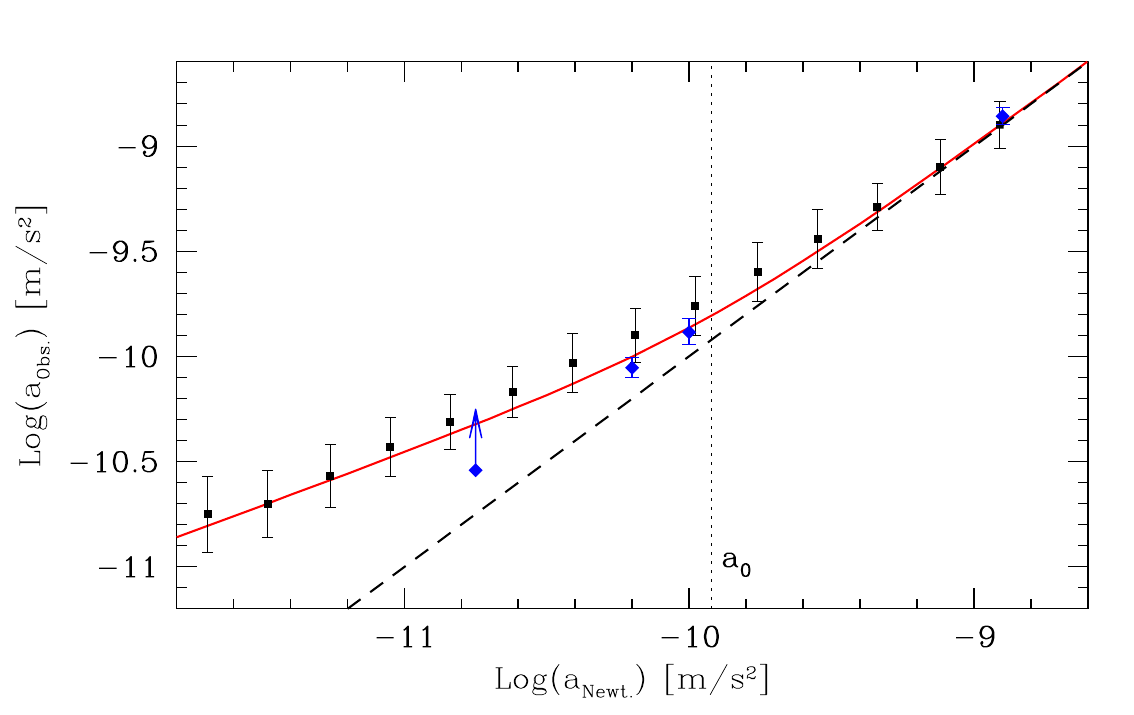}
\caption{ 
Observed (black squares) versus Newtonian acceleration as derived from galaxy rotation curves from binned average of the data by \citet{Lelli17}. The  deviation from the Newtonian value (dashed black line) is clearly apparent. The red line is the Klein interpolation function.  The blue diamonds represent the acceleration probed by wide binary systems from \citet{Chae24} (see text).}
\label{fig:rarbin}
\end{figure}

Wide binary stars (WBSs) have been proposed as a powerful testbed to discriminate between these two interpretations, as they probe the same low-acceleration regime characteristic of galaxies, but on vastly smaller spatial scales where dark matter is not expected to contribute significantly \citep[see, e.g.,][]{Banik}. 
This line of investigation has gained considerable momentum in recent years, largely owing to the high-precision astrometric data provided by the \textit{Gaia} satellite, which have enabled the construction of extensive samples of WBSs with projected separations exceeding $0.025$ pc. Current analyses rely primarily on projected (two-dimensional) kinematics, and the resulting conclusions remain debated. Some studies find consistency with Newtonian expectations \citep[e.g.][]{Pittordis19, Banik}, whereas others report evidence suggestive of a gravitational anomaly in the low-acceleration regime \citep[e.g. ][]{Hernandez23,Chae24}. 

On laboratory scales, measurements of the gravitational constant $G$ exhibit a scatter significantly larger than implied by their quoted uncertainties, a long-standing discrepancy commonly attributed to unidentified systematic errors (see, e.g., \citealt{Xue}).
It has been proposed \citep{Klein} that this dispersion may instead reflect the
same type of deviations from Newtonian dynamics suggested by
astrophysical observations. He showed that agreement could be achieved in terms of MOND, if each experiment experienced a different amount of deviation from Newtonian dynamics depending on the strength of the gravitational field produced by the source masses, as well as, crucially, the relevance electromagnetic and non-inertial forces had in the experiment. The outcome is that after correction all the values from the most recent and precise measurements of $G$ converges toward a single value. This, if confirmed, would provide a significant connection of MOND to laboratory results. 

In this work we take at face value the reported deviations in all three regimes and investigate whether a single MOND-like interpolation function can consistently account for laboratory measurements of $G$, the dynamics of WBSs, and galaxy rotation curves.

\section{MOND basics}

The basic idea of MOND is that below $a_0$ the acceleration field of an isolated point mass deviates from the Newtonian acceleration and should become 
\begin{equation}
a_{\rm mond} = a_N \times F(|a_N|/a_0)
\end{equation}
where $a_N$ is the Newtonian acceleration. 
The choice of the interpolation function $F(a_N/a_0)$ smoothly joining the Newtonian to the MOND regime is not defined except for its asymptotic limits $F(|a_N|/a_0 \rightarrow \infty) = 1$ and $F(|a_N|/a_0 \rightarrow 0) = \sqrt{a_0/a_N}$. The function $F$ can be considered either as a boost factor  of the Newtonian acceleration, or as a boost for $G_N$ ($G = F G_N$).

When applying MOND to laboratory {\bf measurements of $G$, \citet{Klein} adopted} an interpolation function different from those commonly used in astrophysical applications:

\begin{equation}
   F_{\rm Klein}(|a_N|/a_0) = \left[1+\left(\frac{a_0}{|a_N|}\right)^\beta\right]^{\frac{1}{2\beta}} .
\end{equation}

Klein allowed the parameter $\beta$ to vary and found that consistency among the measured values of $G_N$ is achieved only within the narrow interval $1.2<\beta<1.4$. A non-trivial result, since $\beta\simeq1.3$ renders $F_{\rm Klein}$ virtually indistinguishable from the standard MOND interpolation function (e.g. \citealt{Gentile}).
As a consequence, the laboratory tuned $F_{\rm Klein}$ can be used to describe galactic dynamics, but not vice versa, since the two functions differ significantly in the laboratory regime.

From the MOND hypothesis, one can directly derive the \citet{Tully} relation between the asymptotic rotation velocity V$_a$  and the galaxy luminosity L  (i.e., $L \propto V_a^4$), predicted by MOND to become $V_a^4=G M a_0$, where $M$ is the total mass of the galaxy.

\section{Joining the Three Regimes}
\label{join}

\begin{figure}
\centering
\includegraphics[width=1\columnwidth, height=0.85\columnwidth, trim= 80 210 70 200, clip]{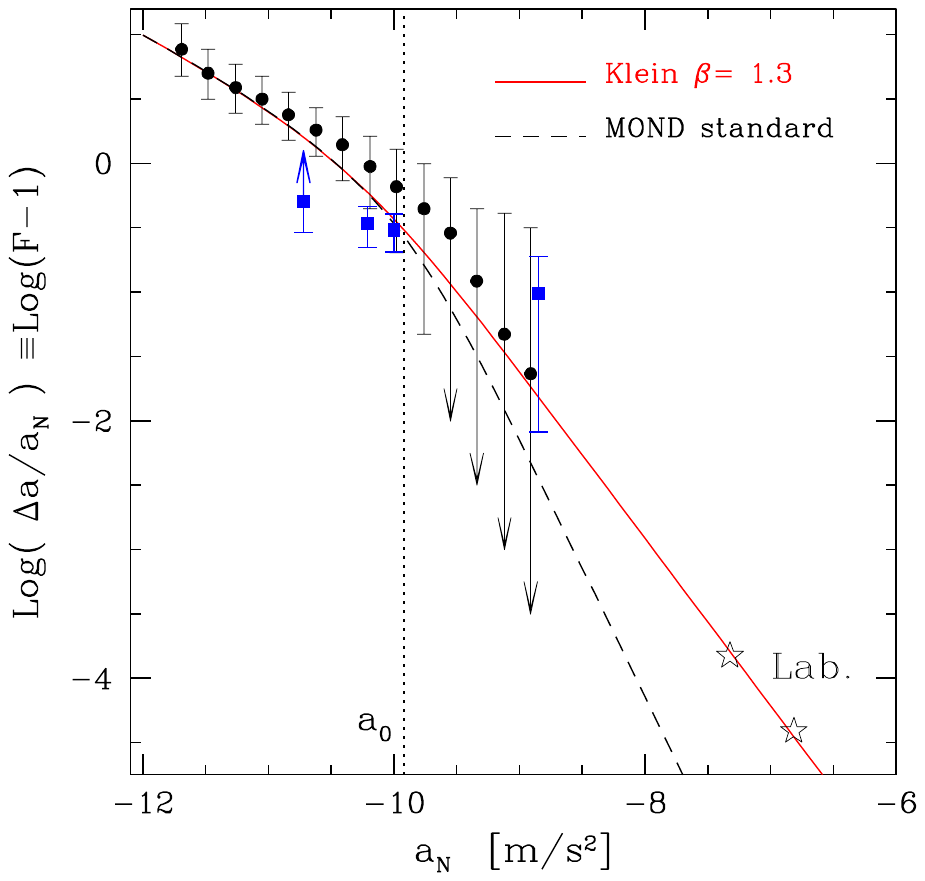}

\caption{Observed $G$ boost, defined as $(a_{\rm obs}-a_{N})/a_{N}$ as a function of the Newtonian acceleration $a_N$.  Points with error bars (1$\sigma$) are from galaxy rotation curves \citep{Lelli17}. Filled squares with error bars are from the WBS analysis by \citet{Chae24}.
Note that value at acceleration $\sim 10^{-11}\ \mathrm{m\,s^{-2}}$ is a lower limit due to  the selection  of WBS (see text). 
The Klein function, calibrated on laboratory measurements (stars), describes galaxies, WBS, and laboratory data. The standard MOND interpolation function is also shown for comparison.
}
\label{fig:kleinBoost}
\end{figure}

In galaxy rotation curves, wide binary stars, and laboratory experiments deviations from Newtonian dynamics when applied to baryonic matter are reported, all indicating an effective increase of $G$.

Here we focus on the results from the Spitzer infrared photometry and accurate rotation curves (SPARC) database \citep{Lelli17}.
The data from galaxy rotation curves cover some 4 orders of magnitude in acceleration. The point of interest is the deviation of the observed radial acceleration $V^2/r$ vs the baryonic Newtonian gravitational acceleration ($GM/r^2$), where $V,r,$ and $M$ are the velocity, the radius, and the baryonic (gas+stars) mass of the galaxies. It is  clear that the  Klein function is a good representation  of the observed data (see Fig. \ref{fig:rarbin}  ). 
Wide binary stars from \textit{Gaia} probe accelerations in the same regime explored by galaxies, and the reported deviations are of comparable magnitude \citep{Chae24}. In particular, measurements around $a_0$ are broadly consistent with the Klein interpolation function (see Fig.~\ref{fig:rarbin} ), as also illustrated by the behaviour of the boost factor $\Delta a/a_N$ as a function of acceleration (Fig. \ref{fig:kleinBoost} ). The point at $a \sim 10^{-11}$ m s$^{-2}$ should be regarded as a lower limit, since the stringent velocity-difference cut adopted in the WBS selection restricts the maximum observable deviation from Newtonian expectations. 
Overall, galaxy rotation curves, WBS data, and laboratory $G$ measurements (open stars in Fig. \ref{fig:kleinBoost}) are all reasonably described by the Klein function.

\section{Discussion and Conclusions}

\begin{figure}
\centering
\includegraphics[width=0.90\columnwidth,height=0.65\columnwidth, trim= 30 150 50 270, clip]{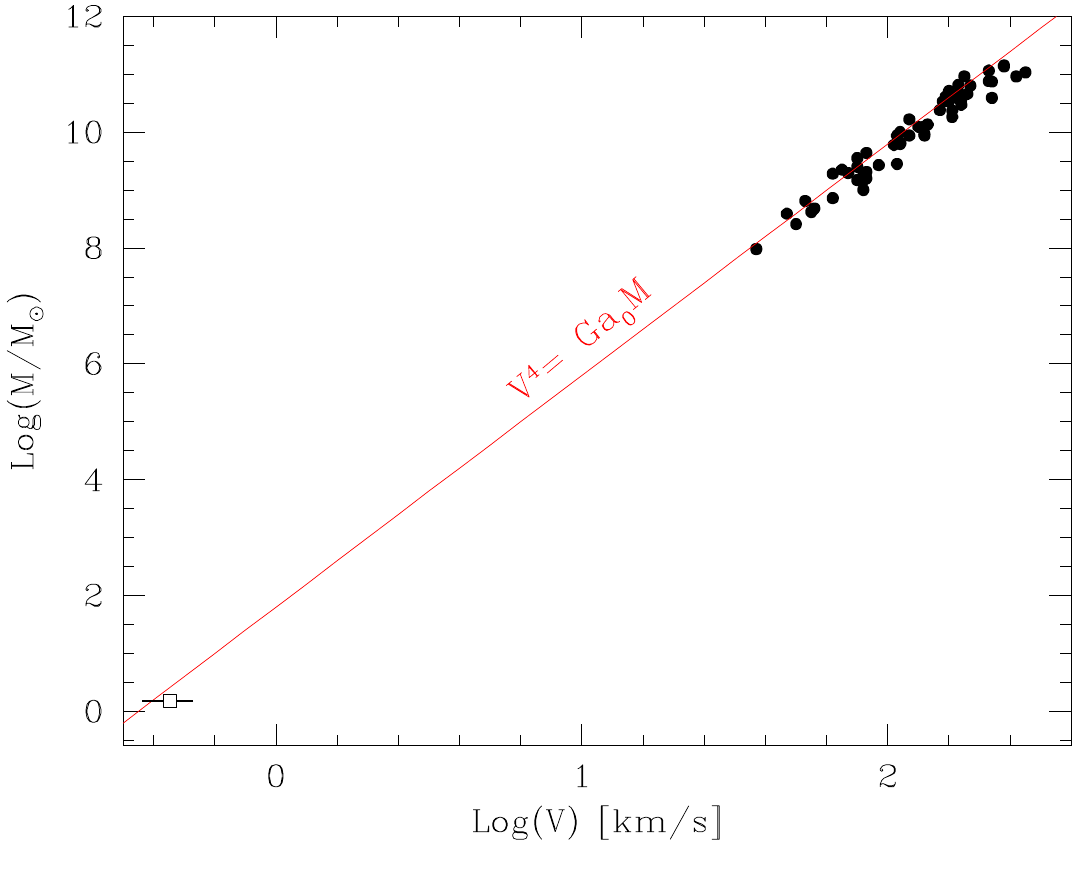}
\caption{The baryonic Tully-Fisher relation for nearby spiral galaxies (filled circles) from the "accurate distance sample" of \citet{Lelli17}, compared to the result for the wide binary stars (open square) from \citet{Hernandez23}. The red line shows the MOND relation $V^4=Ga_0 M$ that describes remarkably very well both observations of galaxies and stars in spite of 10 order of magnitude difference in mass.}
\label{TF}
\end{figure}

While the study of galaxy rotation curves is already a mature field, additional investigation on WBS and dedicated laboratory measurements are urgently required, since in these two cases dark matter cannot play a role. 
In the case of WBS, the available data do not provide compelling evidence in favor of the Newtonian or gravitational anomaly scenarios and should be refined as much as possible in the future.
However, what is most relevant is that WBS do probe the Newtonian dynamics on systems with masses and dimensions vastly different from galaxies. This is made clear when considering the baryonic T-F relation. 
The sample of WBS from \citet{Hernandez23} has an average mass $M=1.5M_\odot$. The 3D asymptotic velocity is 0.45 $\pm$ 0.9 km/s. As illustrated in Fig. \ref{TF}, this is fully consistent with the T-F relation for galaxies, extending it by several orders of magnitude. 
In order to solve the controversy on WBS it is of fundamental importance to extend and refine their study. In particular a key point is to clarify the value of the $G$ boost, at large star separation, corresponding to acceleration $a_N \sim 10^{-11}\ \mathrm{m\,s^{-2}}$ (see e.g. Fig.~\ref{fig:kleinBoost}). 
Help might come from new and more accurate GAIA data (as soon as the DR4 will become available). Most important, the radial velocity of all WBS should be measured with accuracy $\sim 10\ \mathrm{m\,s^{-1}}$ in order to compute the full 3D velocity vector, making possible a complete dynamical analysis of each WBS (see e.g. \citealt{Saglia25}). At present, the available—yet still limited—measurements of the three-dimensional velocities of WBS yield controversial and sometimes conflicting results (see \citealt{Saglia25} and \citealt{Chae26}).

In the case of laboratory experiments,
new measurements of $G$ performed with larger source masses—thereby minimizing possible MOND effects—should yield mutually consistent values, independent of experimental setup. Conversely, experiments deliberately probing lower accelerations could reveal measurable deviations. For example, according to Klein, the deviations in the electrostatic-servo technique {\bf experiment by \citet{Quinn} should} follow exactly the $F$ function. If the experiment were repeated with the same setup but for instance using source masses producing half and twice the original acceleration, the expected value of $G$ would be $6.6769$ and $6.6747$, respectively, assuming $G_N = 6.6742\times 10^{-11}\ \mathrm{m^3\,kg^{-1}\,s^{-2}}$ as proposed by Klein. The resulting $2200$ ppm difference lies within the capabilities of modern experiments. Such tests could offer a direct and potentially decisive probe of gravitation in the low-acceleration regime, with implications extending from the laboratory to galaxies.

The unification scenario that we have presented is essentially based on MOND formula (1). This is obviously a preliminary step, and will require a deeper interpretation. Among the various difficulties, as already pointed by \citet{Klein}, there is the issue of the possible importance of the External Field Effect, raised already in the original Milgrom paper. If it should be accounted for, MOND effect would not be present in lab experiments, because of the Earth field. Note that \citet{Chae24} 
 suggested that his WB data were affected by the Galaxy field. 


While the baryonic T–F relation is firmly established in the local Universe ($z \lesssim 0.1$), where rotation curves are well resolved and baryonic masses can be reliably determined, the situation at higher redshift is considerably less secure. The available samples are smaller and characterized by lower spatial resolution, increased dynamical complexity, and larger systematic uncertainties, so any apparent extension of the relation beyond the nearby Universe should be regarded as provisional.
Moreover, our discussion ignores the relativistic formulation of MOND (e.g. \citealt{Bekenstein}), and possible Cosmological corrections, which may yield modification of the T-F relation (see e.g. \citealt{genzel20}).

The present discussion is necessarily somewhat speculative — built upon the conjecture of a gravitational anomaly in WBS and on Klein’s interpretation of the discrepancies in measurements of $G$ — the primary objective is to promote further experimental investigations within a unified conceptual framework that encompasses galaxy rotation curves, wide binary systems, and laboratory experiments.

\bigskip
{\bf Data Availability:} There are no new data associated with this article.
\section*{Acknowledgments}
We are grateful to the referee for constructive comments.

\end{document}